\documentclass{elsart}
\usepackage{epsfig}

\begin{document}

\begin{frontmatter}

\title{\Large\bf \boldmath The $\sigma $ pole in $J/\psi \to \omega \pi^+ \pi^-$}

\date{}

\maketitle

\begin{center}
\begin{small}
\vspace{0.2cm}

M.~Ablikim$^{1}$, J.~Z.~Bai$^{1}$, Y.~Ban$^{10}$,
J.~G.~Bian$^{1}$, D.~V.~Bugg$^{19}$, X.~Cai$^{1}$, J.~F.~Chang$^{1}$,
H.~F.~Chen$^{16}$, H.~S.~Chen$^{1}$, H.~X.~Chen$^{1}$
% ...
\\(BES Collaboration)\\

\vspace{0.2cm}
\label{att}
$^1$ Institute of High Energy Physics, Beijing 100039, People's Republic of China\\
$^2$ China Center for Advanced Science and Technology(CCAST), Beijing 100080,
People's Republic of China\\
$^3$ Guangxi Normal University, Guilin 541004, People's Republic of China\\
$^4$ Henan Normal University, Xinxiang 453002, People's Republic of China\\
$^5$ Huazhong Normal University, Wuhan 430079, People's Republic of China\\
$^6$ Hunan University, Changsha 410082, People's Republic of China\\
$^7$ Liaoning University, Shenyang 110036, People's Republic of China\\
$^8$ Nanjing Normal University, Nanjing 210097, People's Republic of China\\
$^9$ Nankai University, Tianjin 300071, People's Republic of China\\
$^{10}$ Peking University, Beijing 100871, People's Republic of China\\
$^{11}$ Shandong University, Jinan 250100, People's Republic of China\\
$^{12}$ Shanghai Jiaotong University, Shanghai 200030, People's Republic of China\\
$^{13}$ Sichuan University, Chengdu 610064, People's Republic of China\\
$^{14}$ Tsinghua University, Beijing 100084, People's Republic of China\\
$^{15}$ University of Hawaii, Honolulu, Hawaii 96822\\
$^{16}$ University of Science and Technology of China, Hefei 230026, People's Republic of China\\
$^{17}$ Wuhan University, Wuhan 430072, People's Republic of China\\
$^{18}$ Zhejiang University, Hangzhou 310028, People's Republic of China\\
$^{19}$ Queen Mary, University of London, London E1 4NS, UK \\
\vspace{0.4cm}
$^{\ast}$ Visiting professor to University of Michigan, Ann Arbor, MI 48109 USA \\
$^{\dagger}$ Current address: Purdue University, West Lafayette, Indiana 47907, USA.
\end{small}
\end{center}

\normalsize

\begin{abstract}
  Using a sample of 58 million $J/\psi$ events recorded in the BESII
  detector, the decay $J/\psi \to \omega \pi ^+\pi ^-$ is studied.
  There are conspicuous $\omega f_2(1270)$ and $b_1(1235)\pi$ signals.
  At low $\pi \pi$ mass, a large broad peak due to the $\sigma$ is
  observed, and its pole position is determined to be $(541 \pm 39) $
  - $i$ $(252 \pm 42 )$ MeV from the mean of six analyses.  The
  errors are dominated by the systematic errors.

\vspace{1cm}

\end{abstract}

\end{frontmatter}
\clearpage
\section{Introduction}

Two important aspects of strong interaction physics at low energies
are color confinement and spontaneous breaking of chiral symmetry.
Better experimental knowledge on the $0^{++}$ low energy hadron
spectrum is very important to the understanding of QCD in the
nonperturbative region. For example, the determination of the lowest
lying $0^{++}$ hadron, named the $f_0(600)$ or $\sigma$, will be very
helpful in understanding how QCD realizes chiral symmetry.


There has been evidence for a low mass pole in the early DM2
\cite{dm2} and BESI \cite{bes1} data on $J/\psi \to \omega \pi ^+ \pi
^-$.  A huge event concentration in the $I=0$ S-wave $\pi \pi$ channel
was seen in the region of $m_{\pi \pi}$ around 500-600 MeV in a $pp$
central production experiment \cite{pp}. This peak is too large to be
explained as background \cite{ishida}.  There have been many studies
on the possible resonance structure in $\pi \pi$ elastic scattering.
A summary of these studies up to 1999 is given by Markushin and Locher
\cite{locher}. It was later proved that the $\sigma$ resonance is
unavoidable in chiral perturbation theory in order to explain the $\pi
\pi$ scattering phase shift data \cite{zheng1}.  An analysis based on
chiral symmetry and the Roy equations has been made by Colangelo,
Gasser and Leutwyler \cite{cpt}; a light and broad resonance was found
with the result $M-i\Gamma/2$ = $(470 \pm 30) - i(295 \pm 20)$ MeV for
the pole position.  Renewed experimental interest arises from E791
data on $D^+ \to \pi ^+ \pi ^- \pi ^+$ \cite{e791}; they find $M=478
^{+24}_{-23} \pm 17$ MeV, $\Gamma = 324 ^{+42}_{-40} \pm 21$ MeV.

Results on $J/\psi \to \omega \pi ^+ \pi ^-$ from $5.8 \times 10^7
J/\psi $ events collected with the upgraded BES (BESII) detector are
presented in this paper.  The upgrade of the BES detector included a
new main drift chamber (MDC) and a new time-of-flight (TOF) system. A
detailed description of the BESII detector is given in Ref.
\cite{bes}.  It has a cylindrical geometry around the beam axis.
Trajectories of charged particles are measured in a vertex chamber
(VC) and the main drift chamber (MDC); these are surrounded by a
solenoidal magnet providing a field of 0.4T.  Photons are detected in a
lead-gas Barrel Shower Counter (BSC). Particle identification is
accomplished using time-of-flight (TOF) information from the TOF
scintillator array located immediately outside
the MDC and the $dE/dx$ information in the MDC.

Partial wave analyses (PWA) are performed on this channel using two
methods.  In the first method, the whole mass region of $M_{\pi^+
  \pi^-}$ which recoils against the $\omega$ is analyzed, the $\omega$
decay information is used, and the background is subtracted by
sideband estimation. For the second method, the region $M_{\pi^+
  \pi^-} < 1.5$ GeV is analyzed, and the background is fitted by
$5\pi$ phase space.  In both methods, different parametrizations of
the $\sigma$ pole are also studied.

\section{Event selection}

Here, the selection of $\omega \pi ^+ \pi ^-$ events is outlined,
where the $\omega$ is observed in its $\pi ^+ \pi ^- \pi ^0$ decay
mode.  Candidate tracks are required to have a good track fit with the
point of closest approach of the track to the beam axis being within
the interaction region of 2 cm in $\sqrt{x^2+y^2}$ and $\pm$20 cm in
$Z$ (the beam direction), polar angles $\theta$ satisfying $|\cos
\theta | < 0.80$, and transverse momenta $> 60$ MeV/c. Photons are
required to be isolated from charged tracks and to come from the
interaction point.  Any photon with deposited energy lower than 30 MeV
in the BSC is rejected. Events are required to have four good charged
tracks with total charge zero and more than one good photon.  The TOF
and $dE/dx$ information are used to identify pions; they largely
reject kaons from background reactions such as $K^+K^-\pi
^+\pi^-\pi^0$.

A four constraint (4C) kinematic fit is applied under the $\pi ^+ \pi
^- \pi ^+ \pi ^- \gamma \gamma$ hypothesis, and $\chi^2_{4C} <40$ is
required.  Events with a $2\gamma$ invariant mass $|M_{\gamma \gamma }
- M_{\pi ^0}| < 40$ MeV/$c^2$ are fitted with a 5C kinematic fit to
$\pi ^+\pi ^-\pi ^+\pi ^- \pi ^0$ with the two photons being
constrained to the $\pi^0$ mass. Events with $\chi^2_{5C} <40$ are
selected.  The resulting $\pi ^+\pi ^- \pi ^0$ mass distribution 
is shown in Fig.  1(a).  If there is more than one mass combination
satisfying requirements, the one closest to the $\omega$ mass is
plotted. The $\omega$ signal is selected by requiring
$|M_{\pi ^+ \pi ^- \pi ^0} - M_{\omega }| \le 40$ MeV/$c^2$.
Separation of $\omega \pi \pi$ from $\omega KK$ is very clean, since
the kinematics of these two processes differ strongly.

\begin{figure}[htbp]
\begin{center}
\includegraphics[width=14.0cm]{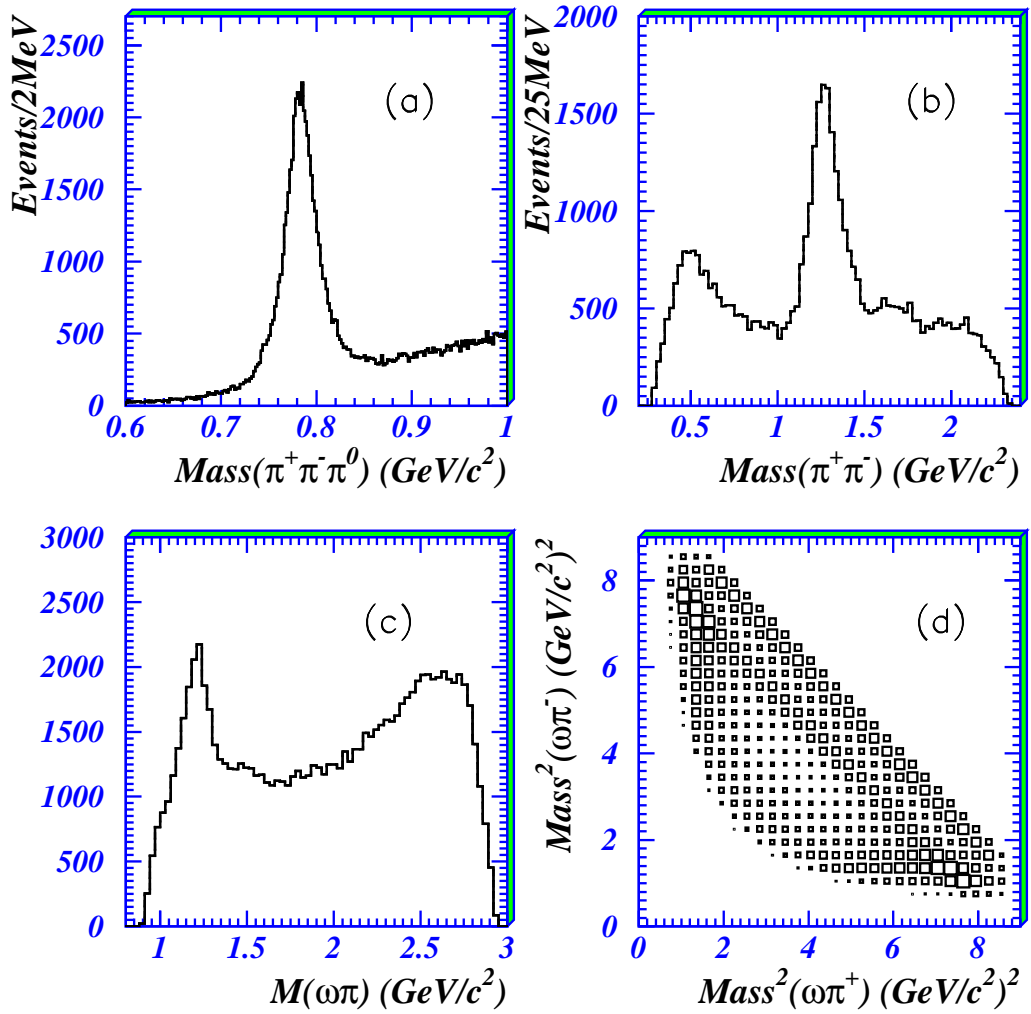}
\caption[]{(a.) Distribution of $\pi ^+\pi ^- \pi ^0$ mass.
  (b.) Distribution of the $\pi^+ \pi^-$ invariant mass recoiling
  against the $\omega$. (c.) Distribution of $\omega \pi$ invariant mass.
  (d.) Dalitz plot.}
\end{center}
\end{figure}

Fig. 1(b) shows the $\pi^+ \pi^-$ invariant mass spectrum which
recoils against the $\omega$, and Fig. 1(c) shows the $\omega \pi$
invariant mass. The Dalitz plot of this channel is shown in Fig. 1(d).
There is a large $f_2(1270)$ peak in Fig. 1(b) and a strong
$b_1(1235)$ peak in Fig. 1(c).  At low $\pi \pi $ masses in Fig. 1(b),
a broad enhancement which is due to the $\sigma$ pole is clearly seen.
This peak is evident as a strong band along the upper right-hand edge
of the Dalitz plot in Fig. 1(d).  Events produced according to phase
space will be broadly distributed over the whole Dalitz plot region.
Therefore this band does not correspond to phase space.  A detailed
Monte Carlo simulation is performed to study the background.  The main
background comes from $5\pi$ events without an $\omega$ and is determined
to be ($14.4 \pm 1.5$)\% using a sideband estimation.


\section{Partial wave analysis}

PWA analyses are performed. Two methods and different parametrizations
of the $\sigma$ pole are used.

For the first analysis,
Method I,
the whole region of $\pi^+ \pi^-$ mass recoiling against the $\omega$
is considered. The channels fitted to the data are:
\begin{eqnarray}
J/\psi &\to & \omega f_2(1270)\\
       &\to & \omega \sigma \\
       &\to & \omega f_0(980) \\
       &\to & b_1(1235) \pi \\
       &\to & \rho(1450)\pi  \\
       &\to & f_2(1565) \omega \\
       &\to & f_2(2240)\omega .
\end{eqnarray}

Amplitudes are fitted to relativistic tensor expressions which are
documented in Ref. \cite{zou}.
For spin 0 in $\pi \pi$, two transitions from $J/\psi$ are allowed
with orbital angular momenta $\ell = 0$ and 2 in the production
process. For spin 2, there are five amplitudes: one with $\ell = 0$,
three with $\ell = 2$ and one with $\ell = 4$.
In fitting these, Blatt-Weisskopf centrifugal barrier factors are
included with a radius of 0.8 fm.
Each amplitude is fitted with a complex coupling constant $G$,
following the usual isobar model.

The subtraction of background is made using sidebands 80 MeV/$c^2$
wide, centered at $M(\pi ^+ \pi ^- \pi ^0)$  = 622 and 942 MeV/$c^2$.
Signal events in the $\omega$ mass band are given positive weight in
log likelihood and sideband events negative weight; the sideband events
(suitably weighted by $\pi ^+\pi ^-\pi ^0$ phase space) then
effectively cancel background in the data sample. Results are stable
when the position and width of the sidebands are varied.

In the amplitude analysis, information from the $\omega \to \pi ^+ \pi
^- \pi ^0$ decay is included in the tensor expressions.  The $\omega$
polarization is given by $\omega_\alpha = \epsilon _{\alpha \beta
  \gamma \delta }(k^{+})^\beta (k^{-})^\gamma (k^{0})^\delta$, where
$k^+, k^-$ and $k^0$ are the 4-momenta of $\pi^+, \pi^-$ and $\pi^0$,
and $\epsilon$ is the fully antisymmetric unit tensor.
Non-relativistically, this reduces to the vector product
$\vec{k^+}\times \vec{k^-}$, where $\vec{k^+}$ and $\vec{k^-}$ are the
three-momenta of the $\pi ^+$ and the $\pi ^- $ in the rest-frame of
the $\omega$.
The spin of the $\omega$ therefore lies along the normal to the
$\omega \to 3\pi$ decay plane. Angular correlations between this
normal and (a) the beam direction, (b) the $\pi^+ \pi^-$ pair provide a
delicate separation between spins 0 and 2 for the pion pair;  they
also identify different $L$ values in the production process.

In the second analysis, Method II,
the $\pi^+ \pi^-$ pairs recoiling against the $\omega$ with
$M_{\pi^+ \pi^-}<1.5$ GeV/$c^2$ are studied.
The covariant helicity coupling amplitude method
is used \cite{wun} to construct the amplitudes. The $\omega$ decay
information is not included in the analysis. In the $\pi^+ \pi^-$
invariant mass spectrum, the processes of $\omega \sigma$, $\omega f_2(1270)$,
and $\omega f_0(980)$ are included in the fit, and
$b_1(1235) \pi$ is considered in the $\omega \pi$ invariant mass spectrum.
The background is approximated by a non-interfering $5\pi$ phase space.

In both analyses, the $f_0(980)$ is parametrized by the Flatt\' e formula which
is written as:
\begin {equation}
f = \frac {1}{M^2 - s - i(g_1\rho _{\pi \pi }(s) + g_2\rho _{K\bar
K}(s))},
\end {equation}
where $\rho (s) = 2k/\sqrt {s}$ and $k$ is the center of mass
momentum of the $\pi$ or $K$ in the resonance rest frame;
we take $M = 0.970 \pm 0.007$ GeV, $g_1 = 0.138 \pm 0.010$ GeV, and
$g_2/g_1 = 4.45 \pm 0.25$.
The full-width at half maximum is $24 \pm 3$ MeV.

For the $f_2(1270)$, only two of the three possible $\ell = 2$ amplitudes
are significant, and the $\ell = 4$ amplitude is small.
In Method I, the $f_2(1270)$ is optimized as a Breit-Wigner with a 
width proportional to $k^5 B_2$, where $B_2$ is the $L=2$ centrifugal barrier,
while in Method II, a constant width Breit-Wigner is used for the $f_2(1270)$.
The $b_1(1235)$ mass optimizes at $1225 \pm 4$ MeV and $1231 \pm 12$ MeV in 
Methods I and II, respectively. These are consistent with the mean mass
quoted by the Particle Data Group (PDG) \cite{pdg}. The $f_2(1270)$ mass is
set to the PDG value in the final fits.
The ratio of D and S-wave decay amplitudes of the $b_1(1235)$ is
consistent with the PDG value of 0.29 and
is fixed to this value.

We come now to the $\sigma$ pole. Several parametrizations of the
$\sigma$ pole are used in the analyses. The first is a Breit-Wigner
with a constant width:
\begin{eqnarray}
 BW_\sigma & = &\frac{1}{M^2 - s  - iM\Gamma _{const.}},
\end{eqnarray}

Secondly, the form introduced by Zou and Bugg \cite{zoubugg} in
fitting $\pi \pi$ elastic
scattering data, consistent with
Cern-Munich data on $\pi \pi$ elastic scattering \cite{munich} and
with $K_{e4}$ data \cite{ke4}, is used for the $\sigma$ amplitude.

\begin {eqnarray}
f & = & \frac {G_{\sigma}}
{M^2 - s  - iM\Gamma _{tot}(s)}, \\
\Gamma _{tot}(s) &=& g_1\frac {\rho _{\pi \pi }(s)}{\rho _{\pi \pi
}(M^2)} + g_2\frac {\rho _{4\pi } (s)} {\rho _{4\pi }(M^2)}, \\
g_1 &=& f(s)\frac {s - m^2_\pi /2}{M^2 - m^2_\pi /2}\exp [-(s - M^2)/a].
\end {eqnarray}

Here $\rho _{\pi \pi }$ is the usual $\pi \pi$ phase
space $2k/\sqrt {s}$, and $k$ is the momentum in the $\pi \pi$ rest frame.
The form includes explicitly into $\Gamma (s)$
the Adler zero at $s = m^2_\pi/2$; the exponential factor
cuts off the width at large $s$. A revised fit to $\pi \pi$ elastic
data and $K_{e4}$ using this formula is presented in \cite{bugg}.

In Eqn. (12), $f(s) = b_1 + b_2s$, where $b_1$ and $b_2$ are adjusted
to reproduce the scattering length and effective range for
$\pi \pi$ elastic scattering; these are from recent
$K_{e4}$ data of Pislak et al. \cite{ke4}.
In the second term of Eqn. (11), $4\pi$ phase space $\rho _{4\pi }(s)$
is approximated by $\sqrt {(1 - 16\mu ^2/s})/ [1 + \exp (2.8 -
s)/3.5]$, with $s$ in GeV$^2$.
In practice, the $4\pi$ width is
significant only at masses above 1200 MeV and has no effect on the
$\sigma$ pole.

Thirdly, we have also tried fitting the $\sigma$ with:
\begin{equation}
BW_{\sigma} = \frac{1}{m_{\sigma}^2 - s - i \sqrt{s}
\Gamma_{\sigma}(s)},
~~~~~\Gamma_{\sigma}(s)
=\frac{g_{\sigma}^2 \sqrt{\frac{s}{4} - m_{\pi}^2 } }
{8 \pi s},
\label{02}
\end{equation}
which was used by the E791 collaboration \cite{e791}.
For the Breit-Wigner amplitude with an $s$-dependent width shown in
Eqn. (13), 
the pole position $\rm {M}$ does not coincide with the mass $M$ where the phase
shift goes through 90$^\circ$ \cite{bugg}.

Fourth, the form given in Ref. \cite{zheng2}, which removes the
spurious singularity hidden in Eq. (13), 
is used:

\begin{equation}
BW_{\sigma} = \frac{1}{m_{\sigma}^2 - s - i \sqrt{s}
\Gamma_{\sigma}(s)},
~~~~~\Gamma_{\sigma}(s)
= \alpha \sqrt{\frac{s}{4} - m_{\pi}^2 } ,
\label{03}
\end{equation}

The fits are made using the maximum likelihood method. 


\section{Results and discussion}
\subsection{Method I}
In method I, when using Eqns. (10) - (12) to describe the $\sigma$ pole (Fit A),
the optimum
fit is obtained with $M = 0.9264$, $g_2 = 0.0024$, $a =
1.082$, $b_1 = 0.5843$, $b_2 = 1.6663$ (all in units of GeV).
Fig. 2 shows the projection of the fit compared with data.

\begin{figure}[htbp]
\begin{center}
\includegraphics[width=12.0cm,height=10.cm]{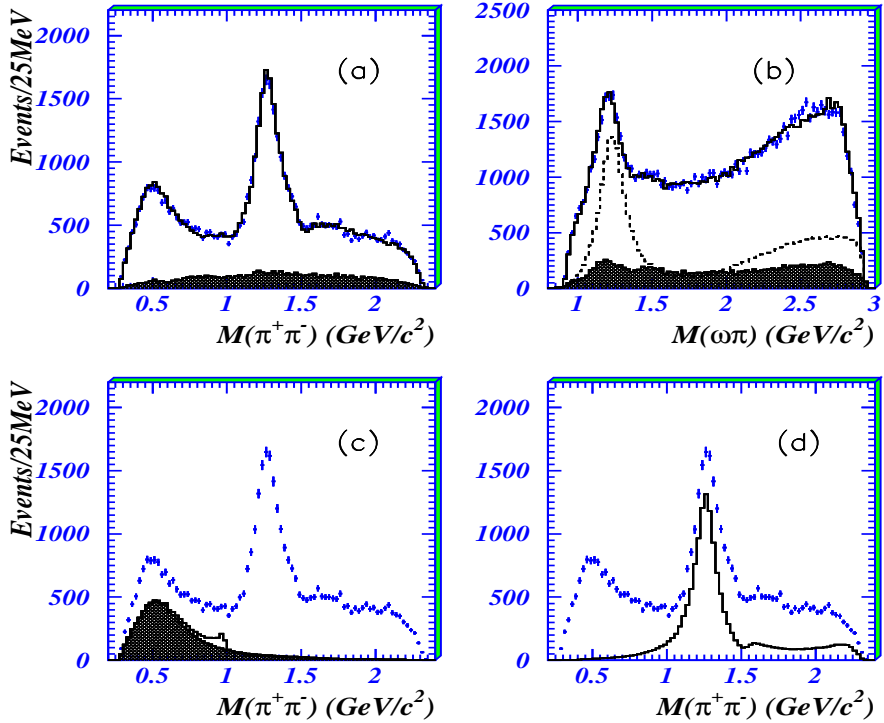}
\caption[]{ The projection of the fit compared with data in Method I, using
            Eqns. (10)-(12) for the $\sigma$ parametrization.
            (a) and (b) show the projections of $\pi \pi$ and
            $\omega \pi$ mass. Histograms show the fit, and the shaded region
            indicates the background estimated from sidebands. The dashed
            curve in (b) shows the fitted $b_1(1235)$ signal (two charge
            combinations). (c) and (d) show the mass projections of
            the $0^{++}$
            and $2^{++}$ contributions to $\pi^+ \pi^-$ from the fit.
            In (c), the shaded area shows the $\sigma$ contribution alone,
            and the full histogram shows the coherent sum of the $\sigma$ and
            $f_0(980)$.}
\end{center}
\end{figure}

The BES data determine the mass and width of the pole well,
but with a rather strong correlation between them.
Fits have been made with 46 variants for $f(s)$ in Eqn. (12).
Optima lie roughly along a line from M = $500-i270$ MeV to
$600-i195$ MeV: the fitted width goes down as the mass goes up.
The optimum is at $(542 \pm 7$ (stat) $\pm 15$ (sys) $\pm 30$ (extrap))
- $i(249 \pm 15$ (stat) $\pm 20$ (sys) $\pm 30$ (extrap))  MeV.
Systematic errors arise roughly equally from
(i) varying the choice of
sidebins,
(ii) varying the magnitude of the background under the $\omega$ peak
of Fig. 1(a), and
(iii) varying the choice of small components in the fit (discussed
below). The last error accounts for systematic errors in the
extrapolation to the pole. This has been estimated from the 46 variants
used to fit BESII data and therefore covers the error due to different
formulae. The systematic errors dominate.

The mass and width of the $f_2(1270)$ optimize at $1271 \pm 5$ MeV and 
$174 \pm 10$ MeV, respectively.
The fitted width of the $b_1(1235)$ is $195 \pm 20$ MeV;
this is  distinctly larger than the PDG value of $142 \pm 9$ MeV.
A fit with the PDG width is visibly poorer, and the log likelihood is worse
than the optimum fit by 302, an enormous amount.
An adequate fit requires a $b_1$ width of at least 180 MeV to
reproduce Fig. 2(b).
DM2 found a similar result \cite{dm2}.
However, changing the width to 142 MeV has
almost no effect on the parameters fitted to the $\sigma$ amplitude.
The reason for this is as follows.
The $\sigma$ band crosses the two $b_1(1235)$ signals on the
Dalitz plot.
The fit effectively integrates over the whole $b_1$ band,
and interferences with the $\sigma$ are affected little by
the precise line-shape of the $b_1(1235)$.

When using the parametrization of Eqn. (13) \cite{e791} for 
the $\sigma$ pole (Fit B), we find an optimum at
$M  = 526 \pm 15$ MeV, $\Gamma _0= 535 \pm 50$ MeV.
This gives a pole position of  $M = (570 \pm 7$ (stat) $\pm 19$ (sys)) -
$i(274 \pm 14$ (stat) $\pm 22$ (sys)) MeV, in satisfactory agreement with
Fit A.

The Breit-Wigner amplitude of constant width for the $\sigma$ given in
Eqn. (9)
gives a very similar intensity distribution to that of Fig. 2.
The mass and width optimize at $M = 470 \pm 20 $
MeV, $\Gamma = 613 \pm 60   $ MeV.
Again the mass and width are strongly correlated; the errors cover these
correlations. Table 1 summarizes the $\sigma$ pole positions obtained
with Method I. In all cases, convergence of the fit is rapid, and the
solution is unique.

\begin {table}[htp]
\begin {center}
\begin {tabular} {|cc|}
\hline
BW Function & Pole position (MeV)\\\hline
Eqns. (10)-(12) & $542 \pm 7 \pm 15 \pm 30$ (extrap)  -
$i(249 \pm 15 \pm 20 \pm 30 $ (extrap)) \\
 Eqn. (13)  & $570 \pm 7\pm 19 
 - i(274 \pm 14 \pm 22 ) $ \\
 Eqn. (9) & $542 \pm 7\pm 20 
- i(269 \pm 15 \pm 25) $ \\
\hline
\end
{tabular}
\caption{Pole positions of the $\sigma$ for three fits from
Method I. Here, the first error is statistical, the second is
systematic
and the last error (30 (extrap) MeV) accounts for the errors
in the extrapolation to the pole, which can also be applied 
to other two fits in the table.}

\end {center}
\end {table}


Angular distributions are shown in Fig. 3 for different
slices of $\pi \pi$ mass.
The third distribution, $\cos \alpha _\pi$ departs
significantly from isotropy. This effect was observed
in the earlier DM2 data [1].
Up to $M(\pi \pi) = 800$ MeV, the departure from isotropy
is due entirely to interference with the $b_1(1235)$.

\begin{figure}[htbp]
\includegraphics[width=14cm,height=8.cm]{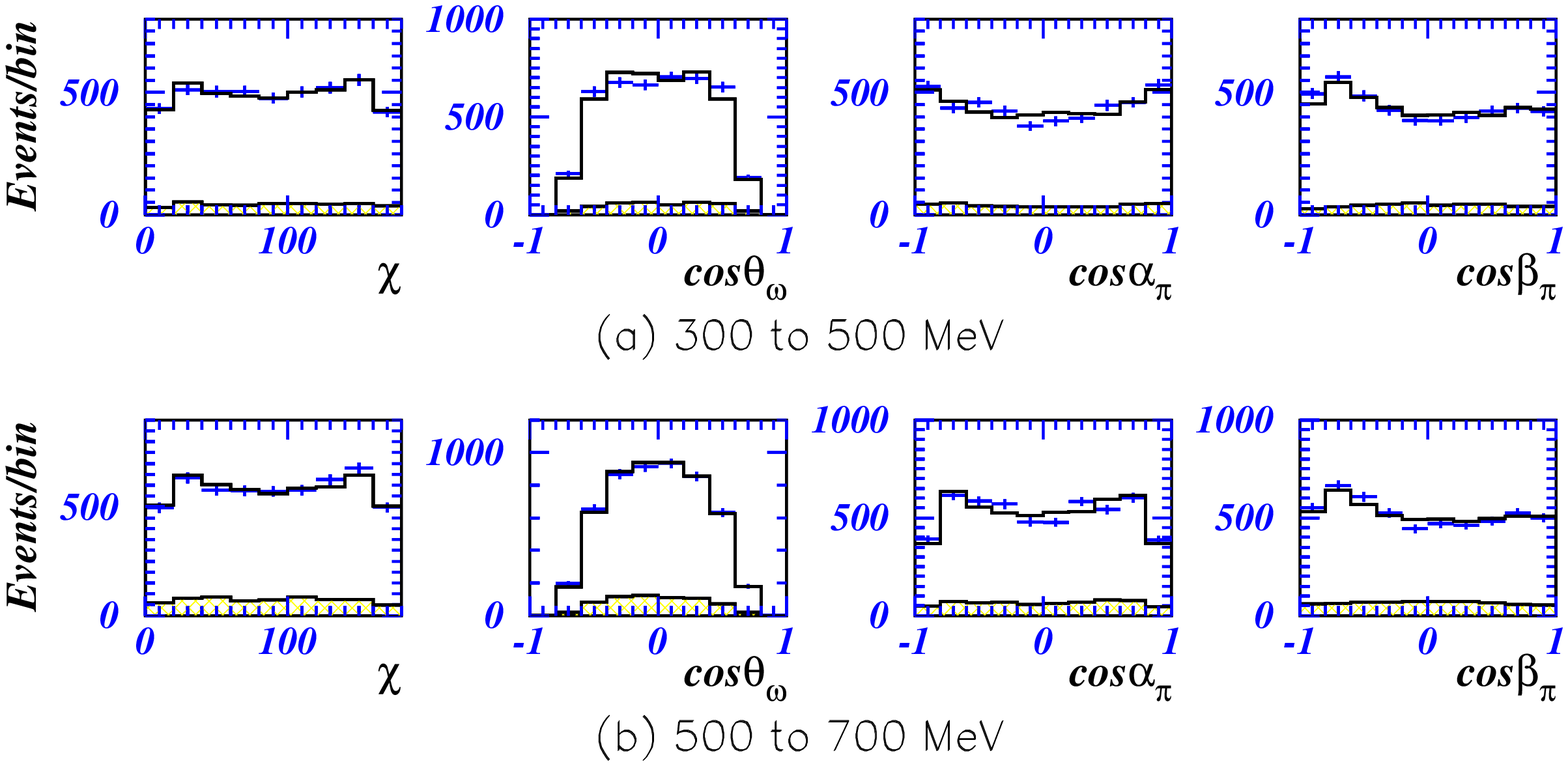}
  \caption[]{Angular distributions for angles $\chi$, $\theta
             _\omega$, $\alpha _\pi$ and $\beta _\pi$ in Method I when
             Eqns. (10)-(12) are used for the $\sigma$ parametrization. Here,
             $\chi$ is the azimuthal angle between the
             production plane of $J/\psi \to \omega X$ and
             the decay plane $X \to \pi \pi$, $\theta _\omega$ is 
             the production angle of the $\omega$ in the $J/\psi$ rest
             frame, $\alpha _\pi$ is the decay angle of the
             $\pi ^+$ in the rest frame of $X$, taken with
             respect to the direction of the recoil $\omega$, and
             $\beta _\pi$ is the angle of the $\pi ^+$ with respect 
             to the direction of $X$ in the rest frame of the $\omega$.
             Histograms show the fit. Slices of $\pi \pi$ mass are 
             (a) 300-500 MeV and (b) 500-700 MeV.}
  \label{angl-wpp1}
\end{figure}

The changes in log likelihood after removing or
adding some components in the fit are studied with coupling constants 
of all other contributions being reoptimised. Our definition of log likelihood 
is such that a change of
0.5 corresponds to a one standard deviation change in
one fitted parameter.
Fig. 4 shows the fits when some of these components are removed.
Fig. 4(a) is the fit without $f_0(980)$.
The log likelihood is improved by 224 when $f_0(980)$ is added.
Therefore, $f_0(980)$ is required. It contributes about 1.1\%
of the intensity.
Errors quoted above for the $\sigma$ pole cover the changes when the
$f_0(980)$ is removed.
A fit without $f_2(1565)$ is shown in Fig. 4(b);
the log likelihood is worse by 210. This fit fails to
reproduce the dip in the $\pi\pi$ mass spectrum at 1560 MeV.
Alternative fits using $f_0(1500)$ fail to
reproduce this feature. Therefore, a small but definite
contribution from $f_2(1565)$ is required with a fitted mass optimising
at 1540 MeV with a width fixed at 126 MeV.

A $\pi ^+\pi ^-$ contribution in the mass range 2000-2250 MeV is
needed. Spin 2 gives the best fit, and a good fit can be obtained
using the $f_2(2240)$, listed by the PDG under $f_2(2300)$ and reported
by Crystal Barrel with a width of 240 MeV \cite{bugg}.  However, it
lies right at the top of the available mass range, so this
identification is not unique. The fit with $f_2(2240)$ removed is
shown in Fig. 4(c).

A definite contribution is required from $\rho (1450)$.
Fig. 4(d) shows the fit omitting this amplitude; it makes the log
likelihood value worse by 923.
Note that there are contributions to this mass projection from
two $\omega \pi$ contributions; this is why removing the
$\rho (1450)$ affects also the high mass region.

We have tried adding a further broad contribution to the $\pi \pi$
S-wave with mass 1500--1600 MeV and a large width of order 800 MeV;
such a contribution has been found by Anisovich et al. in fitting other
data \cite{anisovich}. This gives no significant improvement.
If $f_0(1500)$, $f_0(1710)$ or $f_0(2100)$ are added,
there is no optimum if their masses are scanned.
Fitted contributions are 0.43\% for $\omega f_0(1710)$,
1.1\% for $\omega f_0(1500)$, and $0.36\%$ for $f_0(2100)$.
We regard these as upper limits and omit them from the final fit.

\begin {figure}[htp]
\begin {center}
\epsfig {file=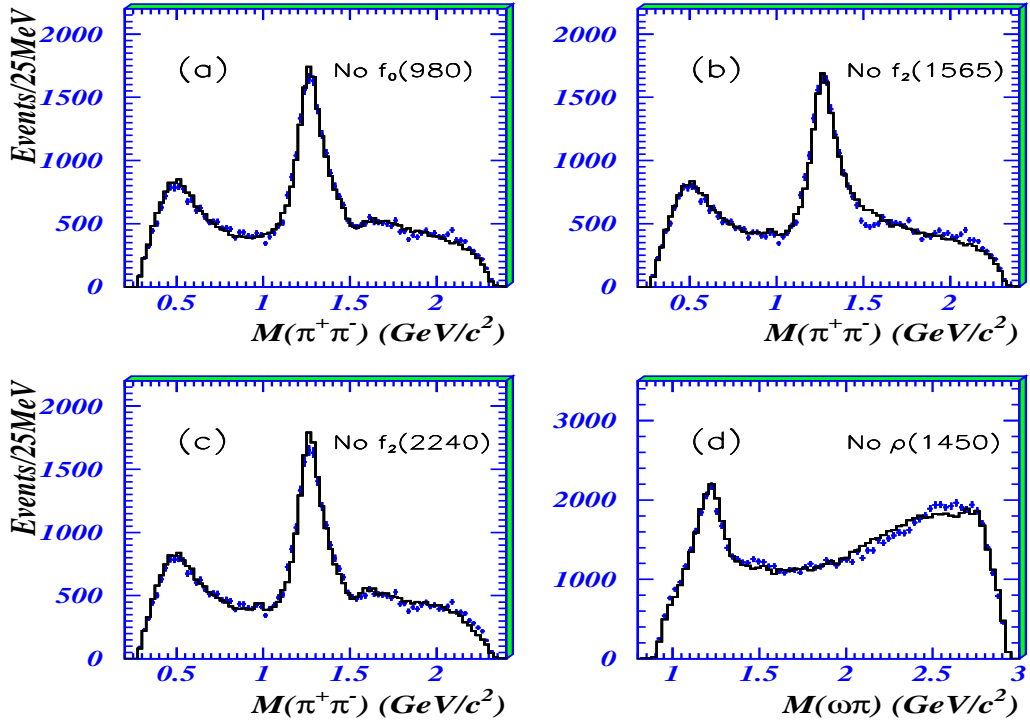,width = 14cm,height=10.cm}
\caption{Method I: (a) Fit without $f_0(980)$, (b) Fit without
$f_2(1565)$, (c) Fit without $f_2(2240)$, and (d) Fit without
$\rho(1450)$.}
\end {center}
\end {figure}


\subsection{Method II}

The second independent analysis, Method II, described in Section 2, is
now reported. Three parametrizations of the $\sigma$ pole -- Eqn. (9),
Eqn. (13),
and Eqn. (14) are used.
The background is fitted by a non-interfering $5\pi$ phase space. 
The width of the $\sigma$ particle is the width at its mass,
i.e., $\Gamma_{\sigma}(m_{\sigma})$.
Masses and widths of the $\sigma$ particle are obtained
from the optimisation.

Fig. 5 shows the projection of the fit compared with data when
using a Breit-Wigner amplitude of constant width (Eqn. (9)). 
The mass and width of the $\sigma$ pole optimise at $M_\sigma =
446 ^{+11 +30}_{-9-32}$ MeV and
$\Gamma_\sigma = 578 ^{+36 +114}_{-23 -86}$ MeV.


\begin{figure}
\begin{center}
\includegraphics[width=6cm,height=5cm]{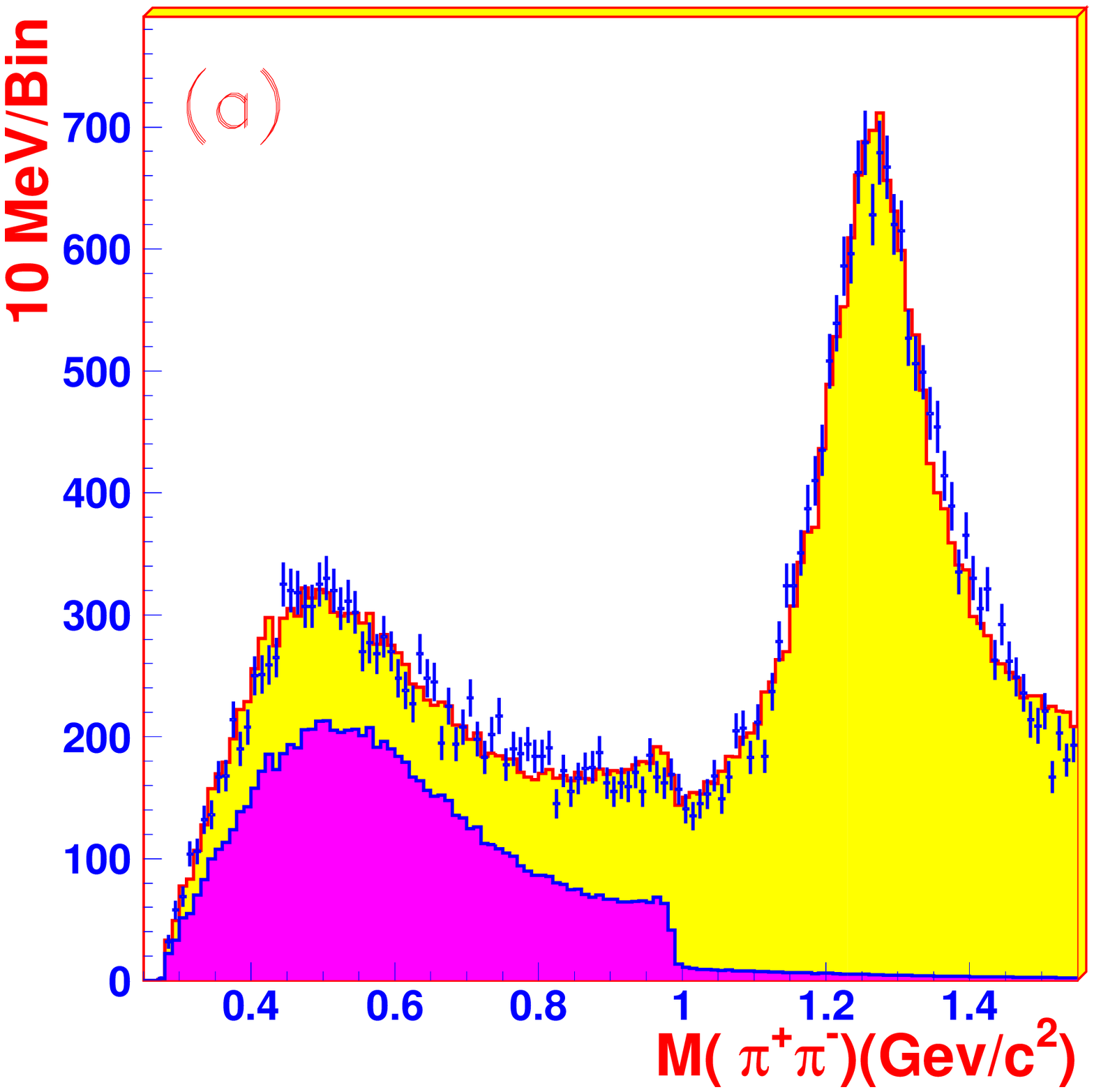}
\includegraphics[width=6cm,height=5cm]{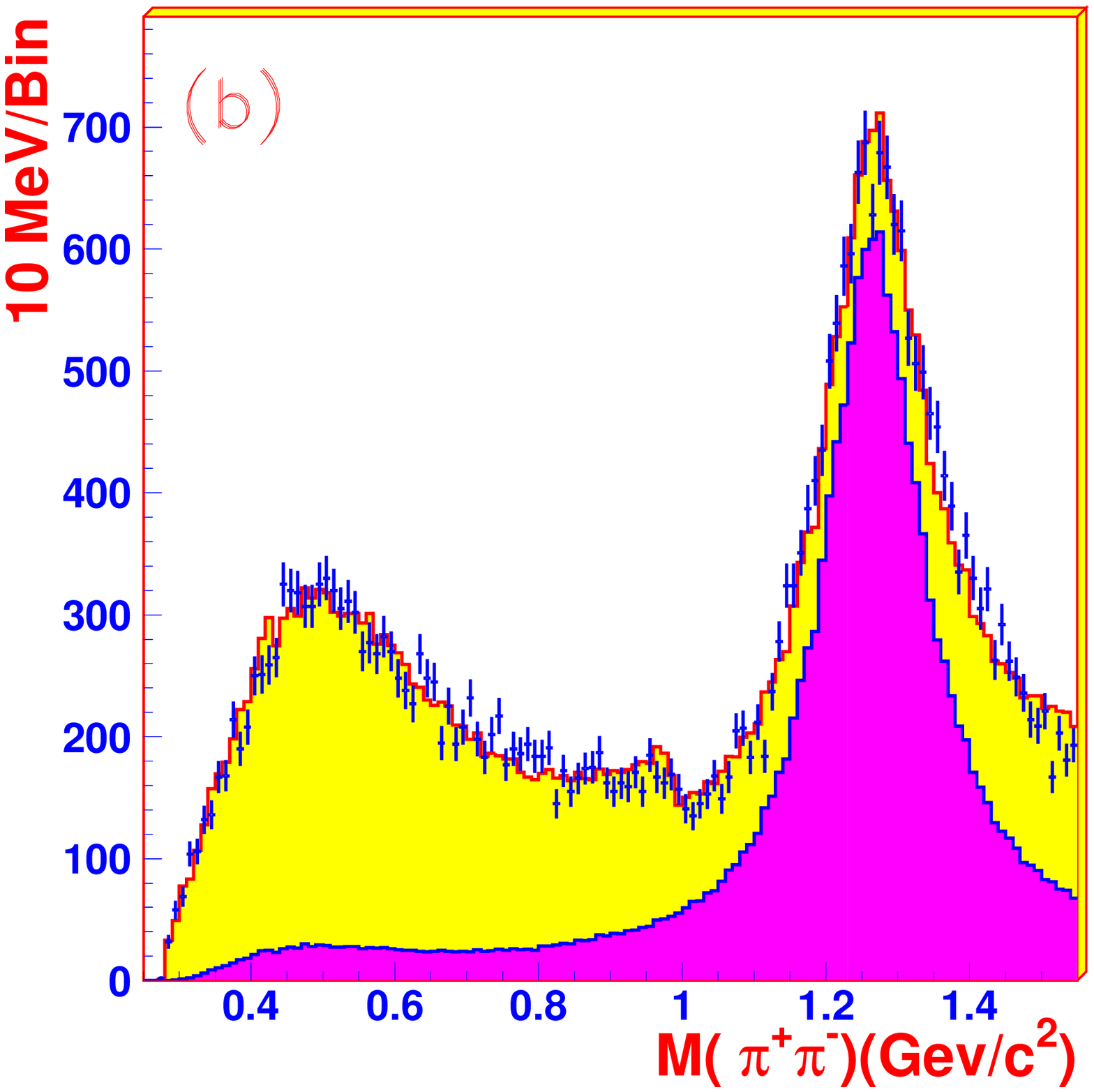}
\includegraphics[width=6cm,height=5cm]{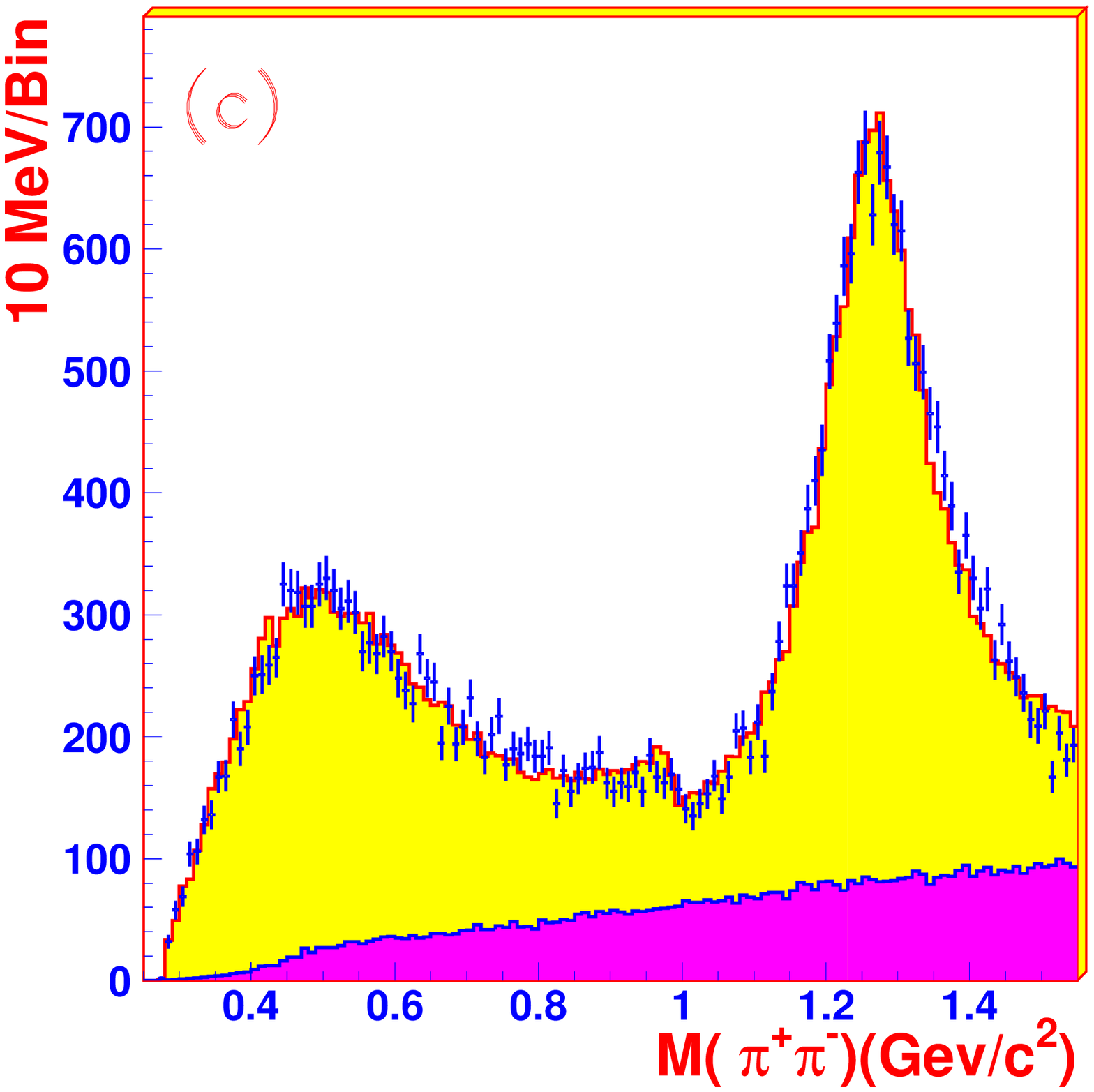}
\includegraphics[width=6cm,height=5cm]{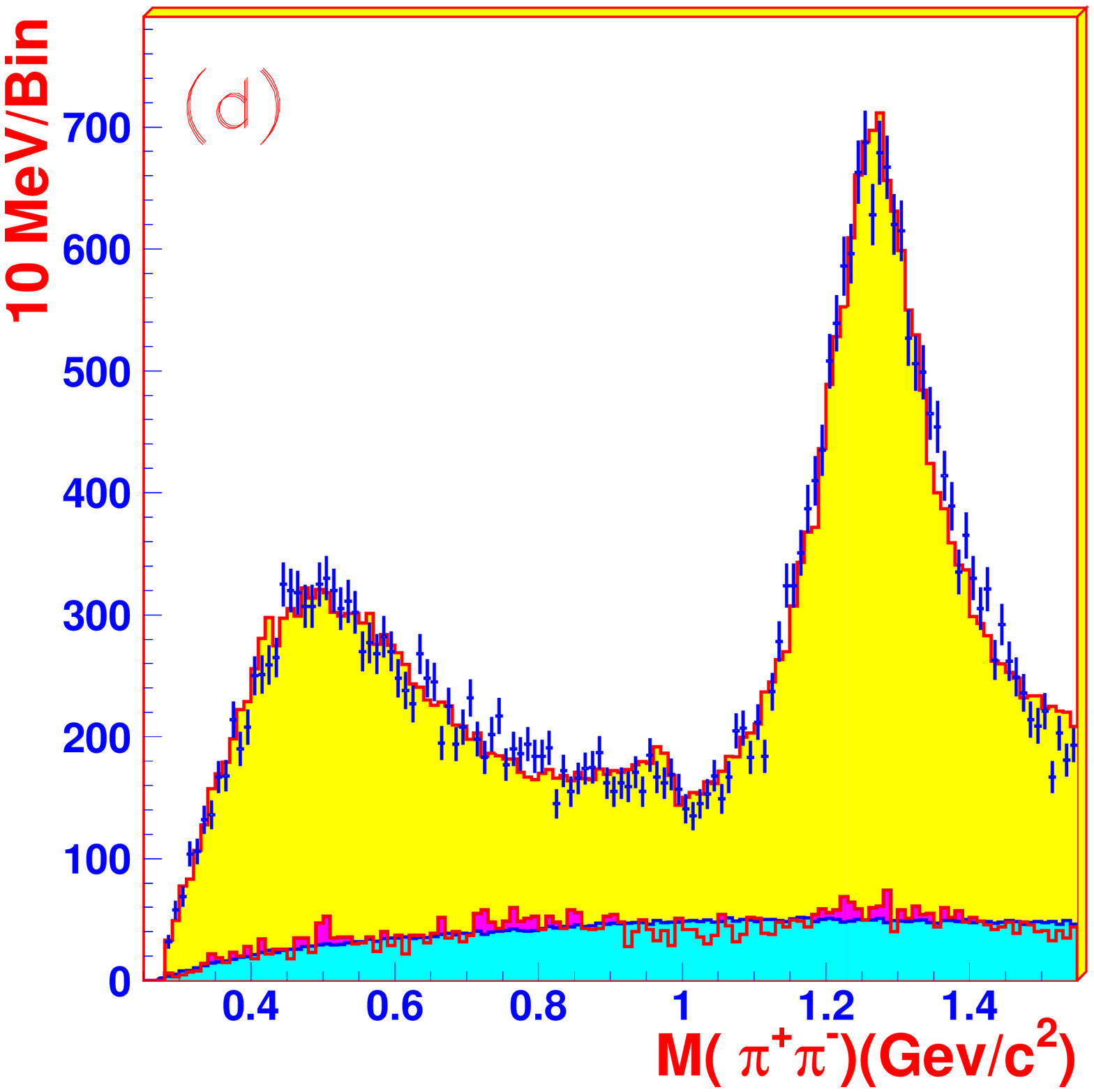}

\caption[]{The projection of the fit in Method II when
          Eqn. (9) is used as the $\sigma$ parametrization compared with data,.
          (a.) The shaded histogram is all $0^{++}$ contributions
                       including the $\sigma$, the $f_0(980)$, and their inteference.
          (b.) The shaded histogram is all $2^{++}$ contributions.
          (c.) The shaded histogram shows the contribution
                       from the $b_1(1235)$.
          (d.) The lower shaded area represents the background, where
               the curve shows the fitted phase space background and 
               the histogram is the background estimated
               from $\omega$ sidebands.}
\end {center}
\end{figure}

The masses, widths, and pole positions of the $\sigma$ are shown in Table
2 when different $\sigma$ parametrizations are used in Method II.  The
first errors are statistical errors, and the second are systematic,
which are determined from the variation for different treatments of
the background (sideband subtraction or direct fit), the changes in
the solution when adding or removing small components, as well as the
differences when changing the background level.
The final global fit to the angular distributions of $M_{\pi^+ \pi^-} < 1.5$
GeV/$c^2$ mass region is shown in Fig. 6. 
The fit agrees well with the data.


\begin{table}[htp]
\begin{center}
\doublerulesep 0pt
\renewcommand\arraystretch{1.5}
\begin{tabular}{|cccc|}
\hline
\hline

BW Function & Mass (MeV)  & width (MeV)  & Pole Postion (MeV) \\
\hline

Eqn. (9)  & $446^{+11 +30}_{-9 -32}$  &
                   $578 ^{+36 +114}_{-23 -86}$ &
($512^{+16 +36}_{-13 -31} $ ) - $i$ (252 $^{+14 +40}_{-9 -33}$) \\
\hline

Eqn. (13)\cite{e791} & $530^{+10 +28}_{-8 -35}$  & $448^{+22 +119}_{-27 -89}$ &
($558^{+14 +42}_{-17 -46} $) - $i$ (231 $^{+12 +58}_{-14 -45}$) \\
\hline

Eqn. (14)\cite{zheng2} & $752^{+10 +76}_{-10 -77}$  & $984^{+36+348}_{-39-258}$&
($521^{+19+44}_{-18-49}$ ) - $i$ (237 $^{+6+33}_{-7-36}$) \\
\hline

\hline
\end {tabular}
\caption{Masses, widths and pole positions of the $\sigma$ particle for Method II.
        The first errors are statistical, and the
        second are systematic. }
\end{center}
\end{table}

\begin{figure}[htbp]
\begin{center}
\includegraphics[width=12.0cm,height=8.cm]{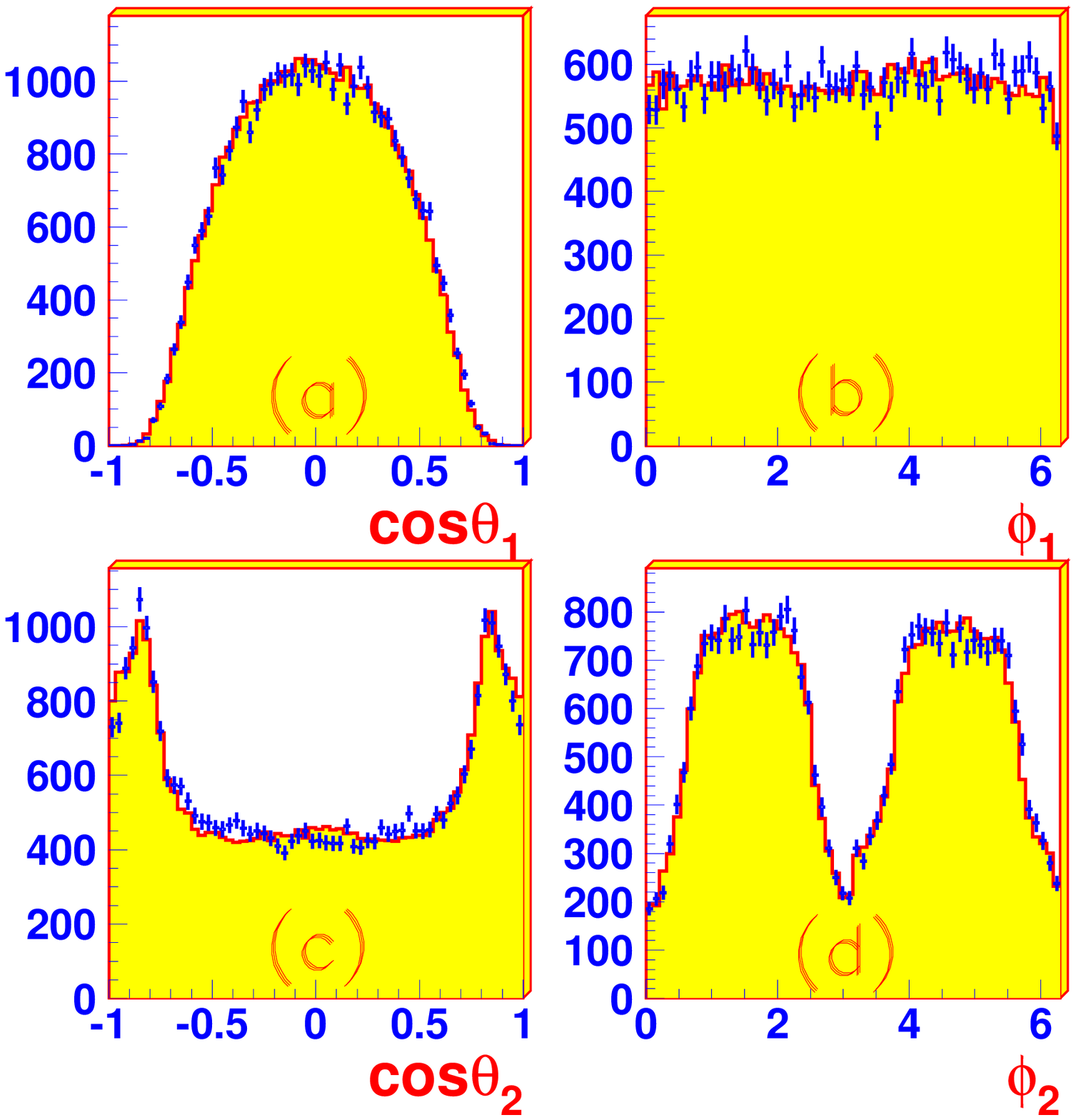}
\caption[]{ The angular distributions in the mass region
$M_{\pi^+ \pi^-} < 1.5$ GeV in Method II when Eqn. (9) is used for
the $\sigma$ parametrization. Here, $\phi_1$ is the
azimuthal angle and $\theta_1$ is the polar angle of $X$ in $J/\psi$ rest
frame for $J/\psi \to \omega X$, and $\phi_2$ and $\theta_2$ are the
corresponding angles of the $\pi^+$ in the $X$ rest frame.
The histogram is the fit, and the
crosses are data.}
\end {center}
\end{figure}

Both the $\sigma$ and the $f_0(980)$ are obviously needed in the data.
Omitting the $\sigma$ makes the log likelihood worse
by 5238. If the $f_0(980)$ is removed, the log likelihood is worse
by 202. The largest resonance in this channel is the $f_2(1270)$
with a mass and width of
$M_{f_2(1270)} = 1268  \pm  4$ MeV and
$\Gamma_{f_2(1270)} = 180 \pm 10$ MeV, respectively.
Another important resonance in this channel is the $b_1(1235)$.
The vertical and horizontal bands in the Dalitz plot (Fig. 1d)
correspond to $J/\psi \to b_1(1235)^{\pm} \pi^{\mp}$.
In the $\pi^+ \pi^-$ mass spectrum,  the $b_1(1235) \pi$
contribution extends
below the $f_2(1270)$. They interfer with the
$\sigma$-particle. The mass and width of the $b_1(1235)$ are 
$M = 1231 \pm 12$ MeV and $\Gamma = 244 \pm 24$ MeV, respectively. As 
in Method I, the width of the $b_1(1235)$ is much larger than 
that of the PDG. However, fixing the width of the $b_1(1235)$ to the PDG value 
does not  change the parameters of the $\sigma$ appreciably. The contribution of the
background is shown in Fig. 5(d).

\section{Summary}

In summary, the essential features to emerge from these data
are the existence of the $\sigma$ and the determination of the $\sigma$ pole position.
Two independent analyses are performed, and different parametrizations of
the $\sigma$ pole are applied.
The mass and width of
the $\sigma$ are
different when using different $\sigma$ parametrizations. However, the pole
position of the $\sigma$ is stable. Different analysis methods
and different parametrizations of the $\sigma$ amplitude give consistent
results for the $\sigma$ pole.
From a simple mean of the six analyses,
the pole position of the $\sigma$ is determined to be
$(541 \pm 39$ - $i$ $(252 \pm 42))$ MeV. Here, the errors from the 
constant width Breit-Wigner parametrisation in Method II are chosen,
which are larger than the errors of the fitted results in Method I.
The systematic errors dominate.

{\vspace{0.8cm}
\vspace{0.4cm}

   The BES collaboration thanks the staff of BEPC for their hard efforts.
This work is supported in part by the National Natural Science Foundation
of China under contracts Nos. 19991480,10225524,10225525, the Chinese
Academy
of Sciences under contract No. KJ 95T-03, the 100 Talents Program of CAS
under Contract Nos. U-11, U-24, U-25, and the Knowledge Innovation Project
of
CAS under Contract Nos. U-602, U-34(IHEP); by the National Natural Science
Foundation of China under Contract No.10175060(USTC),
No.10225522 (Tsinghua University); the Department
of Energy under Contract No.DE-FG03-94ER40833 (U Hawaii),
and the Royal Society
(Queen Mary, London).

One of the authors would like to thank Profs. K. Takamatsu,
S.Ihida, T.Tsuru and M.Ishida for the helpful discussions.


\begin {thebibliography}{99}
\bibitem{dm2} J.E. Augustin et al., Nucl. Phys. B320 (1989) 1.
\bibitem{bes1} Ning Wu (BES Collaboration), Proceedings of the XXXVIth
Rencontres de Moriond, Les Arcs, France, March 17-24, 2001.
\bibitem{pp} D. Alde et al., Phys. Lett. B397 (1997) 350
\bibitem{ishida} T. Ishida et al., Proceedings of Int. Conf. Hadron'95,
Manchester, UK, World Scientific, 1995.
\bibitem{locher} V.E. Markushin and M.P. Locher, Frascati Phys. Ser.
15 (1999) 229.
\bibitem{zheng1} Z. Xiao, H. Q. Zheng, Nucl. Phys. A695 (2001) 273
\bibitem {cpt} G. Colangelo, J. Gasser and H. Leutwyler, Nucl. Phys.
B603 (2001) 125.
\bibitem {e791} E.M. Aitala et al., Phys. Rev. Lett. 86 (2001) 770.
\bibitem {bes} J.Z. Bai et al., Nucl. Instr. Meth. A344 (1994) 319
and A458 (2001) 627
\bibitem {zou} B.S. Zou and D.V. Bugg, Euro. Phys. J A16 (2003) 537.
\bibitem {wun} N. Wu and T.N.Ruan, Commun. Theor. Phys. (Beijing, China) 35
(2001) 547 and 37 (2002) 309
\bibitem {pdg} Particle Data Group (PDG), Phys. Rev. D66 (2002) 010001.
\bibitem {zoubugg} B.S. Zou and D.V. Bugg, Phys. Rev. D48 (1993) R3948.
\bibitem {munich} B.Hyams et at., Nucl. Phys. B64 (1973) 134
\bibitem {ke4} S. Pislak et al., Phys. Rev. Lett. 87 (2001) 221801.
\bibitem {bugg} D.V. Bugg., Phys. Lett. B572 (2003) 1
\bibitem {zheng2} H.Q. Zheng et al., Nucl. Phys. A 733 (2004) 235
\bibitem {anisovich} V.V. Anisovich et al., Phys. of Atomic Nuclei 60 (2000)
1410.

\end{thebibliography}
\end{document}